\documentclass[a4paper,12pt]{article}
\usepackage{cite}
\newcommand{\be}{\begin{equation}}
\newcommand{\ee}{\end{equation}}
\newcommand{\bea}{\begin{eqnarray}}

\newcommand{\eea}{\end{eqnarray}}

\begin{document}

\title{
On Hawking Radiation as Tunneling with Logarithmic Corrections}

\author{A.J.M. Medved$\;^a$ and
Elias C. Vagenas$\;^b$ \\ \\
$\;^a$~School of Mathematics, Statistics and Computer Science \\
Victoria University of Wellington \\
PO Box 600, Wellington, New Zealand \\
E-Mail: joey.medved@mcs.vuw.ac.nz \\ \\
$\;^b$~Nuclear and Particle Physics Section\\
Physics Department\\
University of Athens\\
GR-15771, Athens, Greece\\
E-Mail: evagenas@phys.uoa.gr\\ \\ }

\maketitle
\begin{abstract}

There  has been  recent speculation  that the tunneling paradigm
for Hawking radiation could  ---  after   
quantum-gravitational effects  have suitably been incorporated --- 
provide a  means for resolving
the (black hole) information loss paradox. 
A prospective quantum-gravitational effect   is
the logarithmic-order correction  to the 
Bekenstein--Hawking entropy/area law.
In this letter, it is  demonstrated that, even with the inclusion of the 
logarithmic
correction (or, indeed,  the quantum correction up to any 
perturbative order),  
the tunneling formalism 
is still  
unable to resolve the stated paradox.
Moreover, we go on to show
that the tunneling framework effectively constrains the
coefficient of this logarithmic term 
to be non-negative. Significantly, the latter observation 
implies the necessity for including the  canonical corrections in
the quantum formulation of the black hole entropy.

\end{abstract}

\newpage

\section*{}

\par
The intuitive notion of Hawking radiation \cite{Haw} as a quantum 
tunneling event ({\it e.g.}, \cite{KW,EK,SP})  ---
along with the fundamental  principle of energy conservation ---
provides a meticulous framework 
for incorporating  back-reaction
effects into the  black hole  radiative process. To elaborate,
the positive-energy particle of a spontaneously produced pair
(formed just inside of the black hole horizon) can {\it quantum 
tunnel}
its way out of the horizon  to infinity; with {\it energy conservation}
 then being 
enforced by the negative-energy partner.~\footnote{Note that the same
basic idea applies to a de Sitter cosmological horizon, except
that (in this case) the positive-energy particle will tunnel inwards to
an appropriate de Sitter observer.} 
Even at the tree level ({\it i.e.}, {\it sans} back-reaction effects), this
tunneling paradigm can correctly  reproduce the Hawking temperature for
a myriad of black hole and de Sitter spacetimes; see, for instance,
 \cite{PW,Elias1,Allan1} and 
\cite{Van} for many additional references.  Moreover, 
the inclusion of the back-reaction leads to  the  
distinct quantitative prediction 
of a non-thermal emission spectrum.
Such a spectrum is an intriguing result; being suggestive of
a ``loophole'' in the usual argument for the loss
of black hole information. That is, the so-called (black hole) 
{\it   information loss paradox} \cite{Haw2}  is, as  typical
of many relevant discussions \cite{bhip}, predicated on 
the idea of a purely thermal spectrum.

\par
It is normal to associate lost information with a {\it non-unitary}
 quantum theory.
Hence,  it is
of further significance (with regard to the stated
paradox) that the tunneling paradigm directly implies a unitary
process of black hole evaporation \cite{Parikh1,Parikh2}.
Let us, on this point, be more explicit and consider  one of the key
outcomes of the tunneling methodology: Namely,
the tunneling probability or emission rate can  consistently be cast into
the suggestive form
({\it e.g.}, \cite{PW,MV})
\be 
\Gamma \sim
\exp\left[\Delta S \right] \;.
\label{01}
\ee 
(Here, $S$ denotes the Bekenstein--Hawking
black hole entropy \cite{Bek,Haw} or, more generally,
the horizon entropy in the case of de Sitter spacetimes \cite{GHaw}.)
 That is, the logarithm of this probability
is  ({\it modulo} an irrelevant constant) equal to
the change in the black hole entropy 
during the radiative process. Just such an outcome complies perfectly
with statistical considerations ({\it i.e.}, $\ln\Gamma\sim$ final
density  of states$/$initial density of states) and thus implies a  unitarily
evolving system.

In spite of the non-thermal spectrum and implied  unitarity, it is not at 
all clear 
(as first pointed out by Parikh \cite{Parikh1})  as 
to how the
information might  actually be preserved. The obvious answer --- namely,
in correlations between  emitted particles --- is problematic in that such
correlations  do not appear
to exist.~\footnote{At least  at late times. The tunneling methodology,
as it currently stands,
can not yet address the issue of short-time correlations.}
One can see that this is the case by observing the following 
(see \cite{Parikh1}
and below for details):
\be
\ln\left[\Gamma(E_1)\Gamma(E_2)\right]=\ln\left[\Gamma(E_1+E_2)\right],
\label{02}
\ee
where each probability has been expressed as a function
of the emitted particle energy. To put it another way, 
the  probability
of emitting two particles of energy $E_1$ and $E_2$  is exactly the same
as the probability of emitting just a single particle of
the same total energy; meaning that these probabilities
(and the emitted particles themselves) are then necessarily 
 {\it uncorrelated}.
And so, given that the desired correlations are non-existent to begin with, it
is impractical to suggest 
that these  could serve as a conduit for the information.

On the other hand, a  possible resolution to this paradoxical situation has 
recently
been proposed by Arzano \cite{Arzano}. This author has suggested
that, instead of using the conventional (tree-level)
Bekenstein--Hawking entropy  or~\footnote{A note on conventions: We will be 
setting
all fundamental constants equal to unity
and $A$ denotes the cross-sectional  area of the black hole (or cosmological)
horizon under scrutiny.}
\be
S\;=\;{A\over 4}\;,
\label{03}
\ee
one should rather utilize its {\it quantum-corrected} formulation.
More specifically,  
\be
S_{q}\;=\;{A\over 4}\;+\; \alpha\ln\left[{A\over 4}\right]\;+\;{\rm const.}
\;+\;{\cal O}[A^{-1}]\;,
\label{04}
\ee
where $\alpha$ is a model-dependent (dimensionless) parameter that reflects
our  ignorance of the fundamental
theory of quantum gravity.  That  the quantum-corrected entropy
does take on just such a form  has been demonstrated frequently 
in the literature;
see, for example, \cite{Kaul,Carlip,Gour} and  \cite{list}
for many other pertinent citations. Most notably, loop quantum gravity predicts
(for a Schwarzschild black hole) a {\it microcanonical} contribution to 
$\alpha$ of $-1/2$ 
\cite{DL,Meissner}. Given that the black hole has 
somehow  equilibrated with  a suitable  
heat bath, there would also be an additional {\it canonical} contribution of
{\it at least} $+1/2$ ({\it e.g.}, \cite{Kastrup,Majumdar,Gour2}). 
For a recent discussion on this distinction between microcanonical
and canonical contributions to the logarithmic prefactor,
see \cite{Medved}.

To better understand the underlying premise  
--- and then, alas, the failure --- of the proposed
resolution, let us focus on the simple case of a Schwarzschild black
hole (even though  the formal machinery is much more general).
Firstly,
if a black hole of initial mass $M$ emits a particle of energy $E_1$,  it 
follows
that the associated probability
is given by [{\it cf}, $A(M)=16\pi M^2$ and equation (\ref{01})]
\be 
\ln[\Gamma(M;E_1)] =
S(M-E_1)-S(M) =-8\pi ME_1\left(1-{E_1\over 2M}\right)\;;
\label{05}
\ee 
where we have, for the moment, chosen to ignore
the possibility of quantum corrections.
A second emission of a particle of energy $E_2$ will then occur with a  
probability of
\be 
\ln[\Gamma(M-E_1;E_2)] =
S(M-E_1-E_2)-S(M-E_1) =-8\pi 
(M-E_1)E_2\left(1-{E_2\over 2(M-E_1)}\right)\;.
\label{06}
\ee 
The reader should have no trouble verifying that the sum of
equations (\ref{05}) and (\ref{06}) coincides  exactly
with  the  calculation  of 
$\Gamma(M;E_1+E_2)=S(M-E_1-E_2)-S(M)$; which is really just a 
restatement of equation (\ref{02}).

Now let us, as initially suggested by Arzano \cite{Arzano}, 
repeat the computation for
the quantum-corrected entropy $S_q$ (up to the logarithmic order).
The first emission gives us 
\bea
\ln[\Gamma_q(M;E_1)] &=&
S_q(M-E_1)-S_q(M) \label{07} \\ &=& -8\pi 
ME_1\left(1-{E_1\over 2M}\right) \;+\; 2 \alpha
\ln\left[{M-E_1\over M}\right]\;.
\nonumber
\eea 
Then, the second emission yields the result
\bea 
\ln[\Gamma_q(M-E_1;E_2)] &=&
S_q(M-E_1-E_2)-S_q(M-E_1)
\label{08} \\
 &=& -8\pi 
(M-E_1)E_2\left(1-{E_2\over 2(M-E_1)}\right)
\;+\; 2 \alpha
\ln\left[{M-E_1-E_2\over M-E_1}\right]\;.
\nonumber
\eea

Summing these two revised outcomes, we obtain a (combined) logarithmic term
of simply $2\alpha\ln\left[{M-E_1-E_2\over M}\right]$. It should be reasonably
clear that precisely the same logarithmic term 
arises out of $\Gamma_q(M;E_1+E_2)$; that is,
the probability for a single-particle emission of the same total energy.
Consequently,
the quantum analogue of equation (\ref{02}), or
\be
\ln\left[\Gamma_q(E_1)\Gamma_q(E_2)\right]=\ln\left[\Gamma_q(E_1+E_2)\right]
\;,
\label{09}
\ee
must certainly  be true  (at least) to  logarithmic order. In fact, after 
just a few more iterations,    it is not  difficult 
to convince oneself that this result must be  true up  to {\it any} order
of the power-law expansion implied by  equation (\ref{04}). 
Hence, the  inclusion of such  corrections
is {\it not}  sufficient to account for the late-time correlations.
Which  is to say, even  with the  quantum (gravitationally) corrected 
entropy, 
the tunneling formalism
can  {\it still  not}  provide a mechanism for preserving 
the black hole information. 

Also of interest,  Arzano made  a pertinent   observation  which
 essentially goes as follows\cite{Arzano}: 
The logarithmic-correction term implies that $\Gamma\rightarrow 0$
as $E\rightarrow M$ and, as a consequence, provides a
natural way of suppressing emission as
the energy of the emitted particle approaches the initial mass of the 
black hole.
Such a suppression can indeed occur; however, as has  previously gone
unnoticed,
this can {\it only} be  the case if the parameter $\alpha$ is positive.
This constraint becomes quite evident when equation (\ref{07}) is 
exponentiated
(now dropping the subscript on the particle energy): 
\be
\Gamma_q(M;E) =
\left(1-{E\over M}\right)^{2\alpha}\exp\left[-8\pi 
ME\left(1-{E\over 2M}\right)\right] \;.
\label{001}
\ee
Clearly, the logarithmic correction will suppress
``black hole sized'' emissions  ({\it i.e.}, $E\rightarrow M$) 
when $\alpha >0$  {\it but} will have just  the opposite effect
when $\alpha <0$. That is, a negative value of $\alpha$ will
cause the emission probability to {\it diverge} in this limit!

Actually, this outcome is  not much of a surprise, given that
an evaporating Schwarzschild black hole is (when in isolation)  a highly 
unstable system; with
this instability being a direct consequence of a negative heat capacity.
To achieve stability, it
is necessary to immerse the black hole in a suitable heat bath; allowing
the system  to equilibrate until the temperature of the bath attains
 the same value as that of  
the Hawking radiation. (To realize this type of scenario, 
one could either place
the black hole in a reflective ``box'' or, more pragmatically,
create the effects of such a box by working in an  anti-de Sitter
spacetime.) Notably, the correct description of such a system is provided
by the canonical ensemble. Hence, it should also not come as much of a surprise
(and is quite reassuring) 
that canonical corrections appear to be sufficient to
ensure a strictly non-negative value for 
$\alpha$ \cite{Medved}.

To summarize, we have shown (in the context of the tunneling model
for Hawking radiation) that the logarithmic correction to
the black hole entropy  is unable to  neither  resolve
the information-storage problem nor  suppress black hole sized
emissions in a purely microcanonical framework. 
The best way to understand these outcomes is, in our
opinion, as follows: 
Although  the tunneling paradigm and the quantum-corrected 
entropy can be viewed as manifestations of quantum-gravitational principles,
both of these have (in their current guise) been formulated
at only the level of semi-classical gravity.~\footnote{It is probably relevant
that, as Page has convincingly  argued,
the storage of  information in the correlations could neither
be confirmed nor excluded by a perturbative analysis \cite{Page2}.
Presumably, any semi-classical calculation would fall into this
ambiguous class.}
A truly quantum-gravitational
treatment would, in all likelihood, resolve
these (and many other) issues; but it is probably fair to say that no such  
treatment is promptly
 forthcoming. Nevertheless, even semi-classical gravity
provides much ``food for thought'', and 
we commend Arzano
(in spite of our critical
 observations)
 for initiating a promising direction of investigation.

\section*{Addendum} 

It should be noted that 
M. Arzano has since acknowledged
our findings and will be adjusting the paper \cite{Arzano} accordingly.
Furthermore, we anticipate a collaboration with  M. Arzano  
that encompasses both
the current work and \cite{Arzano}. 

\section*{Acknowledgments}

Research for AJMM is supported  by
the Marsden Fund (c/o the  New Zealand Royal Society)
and by the University Research  Fund (c/o Victoria University).
The work  of ECV is financially supported by the PYTHAGORAS II 
Project ``Symmetries in Quantum and Classical Gravity''
of the Hellenic Ministry of National Education and Religions.



\begin{thebibliography}{99}


\bibitem{Haw} S.W. Hawking, Nature {\bf 248}, 30 (1974); \\
 Commun. Math. Phys. {\bf 43}, 199 (1975).
\bibitem{KW} P. Kraus and F. Wilczek, Nucl. Phys. B {\bf 433}, 403 (1995)
[gr-qc/9408003]; \\
 Nucl. Phys. B {\bf 437},
231 (1995) [hep-th/9411219].
\bibitem{EK} E. Keski-Vakkuri and P. Kraus, Phys. Rev. D {\bf 54},
7407 (1996) [hep-th/9604151].
\bibitem{SP} K. Srinivasan and T. Padmanabhan, Phys. Rev. D {\bf 60}, 024007
[gr-qc/9812028]; \\
S. Shankaranarayanan, K. Srinivasan, and T. Padmanabhan,
Mod. Phys. Lett. A {\bf 16}, 571 (2001) [gr-qc/0007022]; \\ 
S. Shankaranarayanan, T. Padmanabhan and K. Srinivasan, 
Class. Quant. Grav. {\bf 19}, 2671 (2002) [gr-qc/0010042].
\bibitem{PW} M.K. Parikh and F. Wilczek, Phys. Rev. Lett. {\bf 85},
5042 (2000) [hep-th/9907001].
\bibitem{Elias1} E.C. Vagenas, Phys. Lett. B {\bf 503}, 399 (2001)
[hep-th/0012134]; \\
 Mod. Phys. Lett. A {\bf 17}, 609  (2002)
[hep-th/0108147]; \\
 Phys. Lett. B {\bf 533}, 302 (2002)
[hep-th/0109108]; \\ 
 Phys. Lett. B {\bf 559}, 65 (2003)
[hep-th/0209185].
\bibitem{Allan1} A.J.M. Medved, Class. Quant. Grav. {\bf 19}, 589
(2002) [hep-th/0110289]; \\  Phys. Rev. D {\bf 66}, 124009
(2002) [hep-th/0207247].
\bibitem{Van} M. Angheben, M. Nadalini, L. Vanzo and S. Zerbini,
``Hawking Radiation as Tunneling for Extremal and Rotating Black
Holes'', arXiv:hep-th/0503081 (2005).
\bibitem{Haw2} S.W. Hawking, Phys. Rev. D {\bf 14}, 2460 (1976).
\bibitem{bhip} See, for an example discussion, \\ J. Preskill,
``Do Black Holes Destroy Information'',  arXiv:hep-th/9209058 (1992).
\bibitem{Parikh1} M.K. Parikh, ``Energy Conservation and Hawking
Radiation'', arXiv:hep-th/0402166 (2004).
\bibitem{Parikh2} 
M. K. Parikh, Int. J. Mod. Phys. D {\bf 13}, 2351 (2004)
 [hep-th/0405160].
\bibitem{MV} A.J.M. Medved and E.C. Vagenas, ``On Hawking Radiation as
Tunneling with Back-Reaction'', arXiv:gr-qc/0504113 (2005).
\bibitem{Bek} J.D. Bekenstein, Lett. Nuovo. Cim. {\bf 4}, 737 (1972); \\
Phys. Rev. D {\bf 7}, 2333 (1973); \\ 
 Phys. Rev. D {\bf 9}, 3292 (1974).
\bibitem{GHaw} G.W. Gibbons and S. Hawking, Phys. Rev. D {\bf 15}, 2738
(1977).
\bibitem{Arzano} M. Arzano, ``Information leak through the quantum horizon'',
arXiv:hep-th/0504188 (2005).
\bibitem{Kaul} R.K. Kaul and P. Majumdar, Phys. Lett. {\bf 84},
5255 (2000) [gr-qc/0002040].
\bibitem{Carlip} S. Carlip, Class. Quant. Grav. {\bf 17}, 4175 (2000)
[gr-qc/0005017].
\bibitem{Gour} G. Gour, Phys. Rev. D {\bf 66}, 104022
[gr-qc/0210024].
\bibitem{list} D.N. Page, ``Hawking Radiation and Black Hole
Thermodynamics'', arXiv:hep-th/0409024 (2004).
\bibitem{DL} M. Domagala and J. Lewandowski, Class. Quant. Grav.
{\bf 21}, 5233 (2004) [gr-qc/0407051].
\bibitem{Meissner} K.A. Meissner, Class. Quant. Grav. {\bf 21},
 5245 (2004) [gr-qc/0407052].
\bibitem{Kastrup} H.A. Kastrup, Phys. Lett. B {\bf 413}, 267 (1997)
[gr-qc/9707009].
\bibitem{Gour2} G. Gour and A.J.M. Medved, Class. Quant. Grav. {\bf
20}, 3307 (2003) [gr-qc/0305018].
\bibitem{Majumdar} A. Chaterjee and P. Majumdar, Phys. Rev. Lett. {\bf 92},
 141301 (2004) [gr-qc/0309026]; \\ Pramana {\bf 63}, 851 (2004)
[gr-qc/0402061].
\bibitem{Medved} A.J.M. Medved, Class. Quant. Grav. {\bf 22}, 133 (2005)
[gr-qc/0406044]; \\
``A follow-up to 'Does Nature abhor a logarithm?' (and apparently
she dosen't)'' arXiv:gr-qc/0411065 (2004).
\bibitem{Page2} D.N. Page, Phys. Rev. Lett. {\bf 71}, 3743 (1993)
[hep-th/9306083].




\end{thebibliography}
\end{document}